\documentstyle[epsf,11pt]{article}
\begin{document}

\hyphenation{pa-ra-met-ri-sa-tion}

\begin{titlepage}
\begin{flushright}
TPR-96-24
\end{flushright}
\vspace{1.5cm}
\begin{center}
{\Large\bf Space-time Characteristics of the Fireball 
       	from HBT Interferometry${}^\dagger$ }
\end{center}
\vspace{1cm}
\begin{center}
 {\large B. Tom\'a\v{s}ik${}^{a,b}$, U. Heinz${}^{a,c}$,
   U.A. Wiedemann${}^a$ and Wu Y.-F.${}^{a,d}$}
\end{center}
\vspace{0.5cm}
\begin{center}
   {\it ${}^a$    
   Institut f\"ur Theoretische Physik, Universit\"at Regensburg,\\
   D-93040 Regensburg, Germany \\
   ${}^b$ 
   Faculty of Mathematics and Physics, Comenius University, \\
   Mlynsk\'a Dolina, SK-84215 Bratislava, Slovakia \\
   ${}^c$ 
   CERN, Theory Division, CH-1211 Geneve, Switzerland \\
   ${}^d$ 
   Institute of Particle Physics, Hua-Zhong Normal University, \\
   Wuhan, China\\}
\vspace{1.5cm}
{November 28, 1996}
\end{center}
\vspace{1.1cm}
\abstract
{We present the Yano-Koonin-Podgoretski\u\i \
parametrisation of the correlation function. Compared
to the conventionally  
used Cartesian parametrisation, this one
provides more straightforward measurement
of the duration of the 
emission process in the fireball and a clearer signal
of the longitudinal expansion, which is expected
in ulrarelativistic heavy ion collisions.}
\endabstract
\vspace{2.5cm}
{$\dagger$ Talk presented at the Workshop on Heavy Ion Collisions,
       	Sept. 2.--5., 1996, Bratislava, Slovakia}

\end{titlepage}

\section{Introduction}
\label{intro}

The  study of  ultrarelativistic heavy ion 
collisions is motivated by the 
prediction of QCD lattice gauge theory
that hadronic matter undergoes at energy
density of 1 GeV/fm a phase transition
into the 
quark-gluon plasma (QGP).
To determine the energy density obtained 
in heavy ion collisions, a measurement of the 
dimension and lifetime of the fireball is needed. 
However, due to very short
lifetime of the examined object 
it is impossible to use  conventional methods
as e.g. the scattering of external particles. 
The most direct measurement of spatio-temporal
characteristics of the collision region is provided
by Hanbury-Brown/Twiss interferometry, a method 
developed originally in 
radioastronomy in the fifties \cite{HBT}.

In the field of particle physics the idea was first applied
by Goldhaber, Goldhaber, Lee and Pais in 1960 \cite{GGLP}.

In the last years this method has been widely used in
ultrarelativistic heavy ion collisions
to investigate the spatio-temporal characteristics 
of the fireball. 
Important conceptual extensions 
have been connected with this application. This is mapped
by numerous reviews and introductory articles in the 
field \cite{GKW79,BGJ90,Z93,P95,W96}. In this article we
review a new parametrisation of the correlation function,
and point to some of its most important features.
We {\em assume}  complete chaoticity of the source.
This assumption has not been proven yet. However, 
it might be supported
by the successful experimental use of this method. 

In the following Section we start by recalling 
a few basic notions and 
relationships of the HBT interferometry. We pay special 
attention to the problems arising from the on-shell
constraint for the bosons  in the final states and from
the shape of the correlation function, which is very
similar to Gaussian even for many different particle
sources. We recall the way how to handle these problems.
This leads to the parametrisation of the 
correlation function by a Gaussian function in three 
dimensions. Sections \ref{cartes} and \ref{YKP} are 
dedicated to these 
Gaussian parametrisations and to the spatio-temporal
interpretation of their  parameters. As a first step
in Section \ref{cartes}
we shortly recall the basic properties of the 
conventional, so-called Cartesian parametrisation. Then 
we present in Section \ref{YKP} 
the recently derived Yano-Koonin-Podgoretski\u\i \
(YKP)
parametrisation. We introduce its properties, pay particular
attention to
the interpretation of the parameters and we compare 
the quality of the information about the source obtained 
via this parametrisation with the information  from the
Cartesian one. The advantageous features of the
YKP parametrisation reside especially in 
the direct measurement of the
emission duration in the fireball and a
straightforward signal of the longitudinal expansion.
We also comment shortly on the technical 
difficulties which can arise in fitting the data
to this parametrisation.

\section{Correlation basics}
\label{corrf}

We start the brief review of basic theory with
the formula describing the connection between the
source and the observed correlations 
\cite{GKW79,S73,P84,CH94}
\begin{equation}
C(q,K) \simeq \frac
   {{\left | \int d^4x \, e^{i q \cdot x} S(x,K)\right |}^2}
   {{\left |\int d^4x \, S(x,K)\right |}^2} .
\label{1}
\end{equation}
Here $C(q,K)$ is the correlation function depending on the 
variables of average momentum $K = \frac 12 (k_1 + k_2)$ and
the momentum difference $q = k_1 - k_2$. The emission 
function $S(x,p)$
expresses the particle source. It is the phase-space
(Wigner) density describing the probability for the boson 
with four-momentum $p$ to be produced of the space-time 
point $x$ \cite{S73,P84,CH94}.

In the experiment the correlation function is measured, and
one would like to obtain the information about the source,
i.e., to extract $S(x,K)$ from the correlation function
$C(q,K)$. Unfortunately, the measured information 
about $S(xK,)$ is not unambiguous due to the 
on-shell constraint for the particles in the final state. 
The momenta fulfill the relation 
\begin{equation}
q \cdot K = 0 \, 
\label{2}
\end{equation}
which can also be written in the form
\begin{equation}
q^0 =  \beta \cdot q \, , \qquad \beta_i = 
 \frac {K_i}{K_0} \approx \frac {K_i}{E_K} \, .
\label{3}
\end{equation}
Due to this relation, eq.(\ref{1}) is not invertible, because
only three of the four components of the momentum difference
are independent. Hence,  the analysis of correlation data is not
possible in a completely model independent way, and model studies 
have to be employed.

Another point which we want to recall here 
is connected to the fact that the 
correlation function has in the region of small $q$
a Gaussian shape for a very wide
class of source functions \cite{GGLP,Z93}, at least for thermal 
models\footnote{Recently it has been found that the Gaussian
shape of the correlation function is not preserved  if some 
of the bosons
originate from  resonance decays 
\cite{WH96a,WH96b,Bolz93,Schlei96,Ornik96}.}.
Then the emission function can be written in the form
\cite{CSH95b,WSH96,HTWW96a,HTWW96b,tamas}
\begin{equation}
S(x,K) = N({\bf K}) \, S(\bar x({\bf K}),K) \, \exp \left [
  - \frac 12 {\tilde x}^\mu ({\bf K}) B_{\mu\nu}({\bf K})
  {\tilde x}^\nu({\bf K}) \right ] + \delta S(x,K) \, .
\label{4}
\end{equation}
Here ${\tilde x}_\mu$ are the space-time coordinates relative 
to the ``source center'' $\bar x({\bf K})$
\begin{equation}
{\tilde x}^\mu = x^\mu - {\bar x}^\mu({\bf K}), \qquad
{\bar x}^\mu({\bf K}) = \left \langle x^\mu \right \rangle
\label{5}
\end{equation}
with the space-time averages over the source defined as
\begin{equation}
\left \langle f(x) \right \rangle = \frac 
   {\int d^4x \, f(x) \, S(x,K)}
   {\int d^4x \, S(x,K)} \, .
\label{6}
\end{equation}
The statement is that only the second moments given by the
matrix $B_{\mu\nu}$ 
are measured. The first two terms in the Gaussian part
on the r.h.s. of eq.(\ref{4}) are not seen in the 
correlation measurement due to the normalization to the
single-particle spectra in eq.(\ref{1}). The correction to
the Gaussian approximation $\delta S(x,K)$ is neglected as
far as one is fitting the data by a Gaussian.
For the class of models to be discussed below,
non-gaussian effects from $\delta S$ 
have been shown to be small \cite{WSH96}.

Inserting (\ref{4}) into (\ref{1}) and neglecting
$\delta S$ one gets
\begin{equation}
C(q,K) = 1 + e^{- q^\mu q^\nu \, (B^{-1})_{\mu\nu}} \, .
\label{6prime}
\end{equation}
This shows that the inverse of the matrix $B_{\mu\nu}$, 
expressible
as
\begin{equation}
(B^{-1})_{\mu\nu} = \left \langle {\tilde x}_\mu
 {\tilde x}_\nu \right \rangle 
\label{7}
\end{equation}
are measured. However, due to the on-shell constraint,
not all 10 components of this matrix are measured. 
It is possible to measure only 6 combinations of the
matrix elements. In what follows, we will treat the
case of azimuthally symmetric sources. This corresponds 
to central collisions. In this case 7 
non-vanishing matrix elements $B_{\mu\nu}$ 
are needed for the 
full description of the source according to eq.(\ref{4}), 
but the number of
measurable combinations of them is reduced to 4.

The question investigated in this work is how
to choose the Gaussian parametrisation of the 
correlator such that  
the interpretation of the four measured parameters
in terms of the space-time dimensions of the source
is as simple (and model-independent) as possible.

Interpreting the correlation parameters (called
also HBT radii) it is important to realize that
the correlations do not measure the dimensions
of the whole source.
According to eqs.(\ref{6prime}) and (\ref{7})
they are combinations of the space-time
variances of the source and they are 
measuring  only the so-called homogeneity
lengths \cite{CSH95b,WSH96,MS88,CSH95a,AS95}. 
It is intuitively obvious that if
the measurement is focussed on bosons with 
given average momentum only the part of the source
producing these bosons is measured. This is relevant
e.g. in sources with strong expansion. Then a particular 
momentum is not produced by the whole fireball, but
only by a small part of the collision region. 
Bosons with different momenta 
originate
from different parts of the source. Hence the HBT radii can
be momentum dependent, and this dependence can carry 
information about the dynamics of the source.
It is  therefore an object of very intensive 
theoretical \cite{WSH96,MS88,CL96,HTWW96c} 
and experimental \cite{NA44,NA35,Alber95,NA49QM96}
study.

\section{Gaussian parametrisations 1: \newline
 Cartesian parametrisation}
\label{cartes}

In this Section we shortly recall the basic
properties of the Cartesian parametrisation. 
This sets the stage for a comparison to the 
Yano-Koonin-Podgoretski\u\i \ (YKP) parametrisation
in the next Section. 

We use the usual coordinate system with the z-axis
in the direction of the beam and the x-axis parallel to
the transverse component of the average momentum
${\bf K}$. Then the z-axis is called also longitudinal,
the x-axis is labeled as outward and the remaining
y-direction is denoted as side-ward.

Both Cartesian and YKP 
parametrisations are constructed for 
azimuthally symmetric events. As already mentioned,
in this case four fit parameters are needed 
for the Gaussian parametrisation of the correlation 
function. Different parametrisations can be obtained
from eq.(\ref{6prime}) by different
choices for the  three independent components
of the momentum difference $q$ according to eq.(\ref{2}).

In the ``standard'' Cartesian parametrisation \cite{CSH95b,CSH95a}
the three 
independent $q$--components are chosen to be the three
spatial components of the momentum difference. The
parametrisation is given by the following formula
\begin{equation}
C({\bf q},{\bf K}) - 1 = \exp [ - R_s^2({\bf K})  q_s^2  - R_o^2
   ({\bf K})  q_o^2 
   - R_l^2({\bf K})  q_l^2 - 2 \, R_{ol}^2({\bf K}) q_s q_l ] .
\label{8}
\end{equation}
The interpretation of the four HBT radii appearing in this 
parametrisation is most straightforward in the so-called
LCMS (Longitudinal Co-Moving System) frame. This is the 
longitudinally boosted frame in which
the average pair momentum has only a transverse
component, $K_l = 0$. 
The interpretation of the HBT radii in this frame
is follows from the model independent expressions
\cite{CSH95b,HB95}
\begin{eqnarray}
R_s^2 & = & \left \langle {\tilde y}^2 \right \rangle
\label{9a}
\\
R_o^2 & = & \left \langle {({\tilde x} - 
  \beta_\perp {\tilde t})}^2 \right
  \rangle
\label{9b}
\\
R_l^2 & = & \left \langle {\tilde z}^2 \right \rangle \label{9c} \\
R_{ol}^2 & = & \left \langle {({\tilde x} - 
   \beta_\perp {\tilde t})}
   \, {\tilde z} \right \rangle
\label{9d}
\end{eqnarray}

It is clearly seen that these radii mix spatial 
and temporal information about the source. It has been
found, that the emission duration can be measured
by the difference \cite{CP91}
\begin{equation}
R_o^2 - R_s^2 =
  \beta_\perp^2 \left \langle {\tilde t}^2 \right \rangle 
  - 2 \beta_\perp \left \langle \tilde x \tilde t \right \rangle +
  \left \langle \tilde x^2 - \tilde y^2 \right \rangle.
\label{10}
\end{equation}
In the typical case of ultrarelativistic heavy ion collisions 
the last two terms at the r.h.s. of this equation can be
treated as perturbation \cite{HTWW96b} 
and the difference is a measure
for the effective emission duration (the 
so-called lifetime of the source). However, two problems 
are connected with this measurement. The first is caused by
the presence of the pre-factor $\beta_\perp^2$, which 
makes this observable  small in the region of most data 
points. The second one is more serious. In typical measurement
in ultrarelativistic heavy ion collisions $R_o^2$ and
$R_s^2$ are typically bigger than
the expected lifetime. Thus the lifetime
measured in this way is obtained with big 
statistical errors.

\section{Gaussian parametrisations 2: \newline
  the Yano-Koonin-Podgoretski\u\i \ parametrisation}
\label{YKP}

The Yano-Koonin-Podgoretski\u\i \ parametrisation 
\cite{YK78,P83,CNH95} is given by the following 
formula
\begin{equation}
C({\bf q},{\bf K}) - 1 = \exp[- R_\perp^2({\bf K})  q_\perp^2 - 
   R_\parallel^2({\bf K}) 
  ( q_l^2 - (q^0)^2 ) - ( R_0^2({\bf K}) + R_\parallel^2({\bf K}) )
  (q \cdot U({\bf K}) )^2 ]
\label{11}
\end{equation}
In this parametrisation, the three
independent momentum difference
components are: $q_\perp = \sqrt{q_o^2 + q_s^2}$, $q_l$
and $q^0$. The fit parameters are now $R_\perp^2$, 
$R_\parallel^2$, $R_0^2$ and the so-called Yano-Koonin
velocity $v$ appearing here in the four-velocity $U$. This 
four velocity is assumed to have only a longitudinal
component
\begin{equation}
U({\bf K}) = \gamma ({\bf K}) \, ( 1,\, 0, \, , 0, \, v({\bf K})),
  \qquad \gamma({\bf K}) = \frac 1{\sqrt{1 - v^2({\bf K})}}\, .
\label{12}
\end{equation}

The YKP parametrisation is constructed in the way that the 
three radius parameters are invariant under longitudinal
boosts, i.e., in any longitudinally boosted
reference frame the analysis of the data
should give the same results. 

The model independent expressions for the HBT radii have
been found in refs.\cite{HTWW96a,HTWW96b}. They are easiest
written with the help of notational shorthands
\begin{eqnarray}
 \label{13a}
   A &=& \left\langle \left( \tilde t
	 - {\tilde \xi\over \beta_\perp} 
	 \right)^2 \right\rangle \, ,
 \\
 \label{13b}
   B &=&  \left\langle \left( \tilde z
	 - {\beta_l\over \beta_\perp} 
	 \tilde \xi \right)^2 \right\rangle
   \, ,
 \\
 \label{13c}
   C &=& \left\langle \left( \tilde t - 
	       {\tilde \xi\over \beta_\perp} \right)
	       \left( \tilde z - 
	       {\beta_l\over \beta_\perp}
	       \tilde \xi \right) \right\rangle \, .
\end{eqnarray}
Here $\tilde \xi \equiv \tilde x + i \tilde y$.
Furthermore we use that
for azimuthally
symmetric sources $\langle \tilde
y\rangle = \langle \tilde x \tilde y \rangle = 0$ 
and $\langle \tilde \xi^2 \rangle = \langle
\tilde x^2 - \tilde y^2 \rangle$.
Then the HBT radii and the YK velocity 
are given by the following formulae
\begin{eqnarray}
 \label{14a}
   R_{\perp}^2 &=& {\langle{ \tilde{y}^2 }\rangle} \, ,
 \\
 \label{14b}
   R_\parallel^2 &=&  B{-}v C,
 \\
 \label{14c}
   R_0^2 &=& A{-}v C ,
 \\
\label{14d}
   v &=& {A+B\over 2C} \left( 1 - 
   \sqrt{1 - \left({2C\over A+B}\right)^2}
		       \right) \, .
\end{eqnarray}

Since eqs.(\ref{8}) and (\ref{11}) are just different
parametrisations of
the same correlation function, there must exist a simple
connection between the YKP parameters and those of 
the Cartesian 
parametrisation. It is given by  \cite{HTWW96a}
 \begin{eqnarray}
 \label{15a}
   R_s^2 & = & R_\perp^2 \, , \\
 \label{15b}
   R_{\rm diff}^2 \equiv  R_o^2 - R_s^2 & = & \beta_\perp^2 \gamma^2
	     \left( R_0^2 + v^2 R_\parallel^2 \right) \, ,
 \\
 \label{15c}
   R_l^2 &=& \left( 1 - \beta_l^2 \right) R_\parallel^2
	     + \gamma^2 \left( \beta_l-v \right)^2
	     \left( R_0^2 + R_\parallel^2 \right)\, ,
 \\
 \label{15d}
   R_{ol}^2 &=& \beta_\perp \left( -\beta_l R_\parallel^2
	     + \gamma^2 \left( \beta_l-v \right)
	     \left( R_0^2 + R_\parallel^2 \right) \right)\, .
 \end{eqnarray}

Although the three radius parameters are longitudinally
boost-invariant, there is a special frame in which their
interpretation is simplest. The frame is defined by the 
relation $v = 0$ and it is called the Yano-Koonin (YK)
frame. In this frame the formulae (\ref{14a}--\ref{14d})
simplify to
\begin{eqnarray}
R_\perp^2 & = & \left \langle \tilde y^2 \right \rangle  \, ,
\label{16a} \\
R_\parallel^2 & = & \left \langle \tilde z^2 \right \rangle -
  2 \frac {\beta_l}{\beta_\perp} \left \langle \tilde z \tilde x 
  \right \rangle + \frac {\beta_l^2}{\beta_\perp^2} \left \langle
  \tilde x^2 - \tilde y^2 \right \rangle \, ,\label{16b} \\
R_0^2 & = & \left \langle \tilde t^2 \right \rangle -
  \frac {2}{\beta_\perp} \left \langle \tilde t \tilde x 
  \right \rangle + \frac {1}{\beta_\perp^2} \left \langle
  \tilde x^2 - \tilde y^2 \right \rangle \, , \label{16c} 
\end{eqnarray}
where $\tilde x$, $\tilde y$, $\tilde z$, $\tilde t$ are
now measured in the YK frame.
It has been pointed out in \cite{HTWW96a} and shown 
in a detailed 
numerical model study in \cite{HTWW96b} that the last
two terms on the r.h.s. of eqs.(\ref{16b}) and (\ref{16c})
can be considered as a small correction within a class 
of thermal models with Gaussian geometrical density profile. 
Hence, $R_\parallel$ measures (up to the corrections) the
dimension in the longitudinal direction in the YK frame 
and $R_0$ is a
measure for the emission duration in that frame. 
In contrast to
the Cartesian parametrisation, the time variance appears here as
the leading term of the fit parameter $R_0^2$. Thus the problems
with the accuracy arising by the subtraction of two numbers
of the same order are avoided here.

The YK velocity $v$ has been found to coincide up to small
corrections with the longitudinal
velocity of the point of maximal emissivity
for bosons of given momentum. This will be illustrated
in the model study in Subsection \ref{model}. This opens the
possibility to investigate the longitudinal expansion of 
the source \cite{HTWW96a,HTWW96b}.

\subsection{Remarks on the fit procedure}
\label{remarks}

Recently it has been found in the experimental analysis
that fits
with the YKP parametrisation tend to be more unstable
than the corresponding Cartesian fit. 
We do not have the 
solution of this problem, but we want to present 
two ideas which might help to solve it.

The YKP parametrisation is constructed in a little bit
sophisticated and complicated way. 
This provides a relatively simple interpretation of the
fit parameters. On the other side, the fitting procedure 
could be more transparent by the usage of a Gaussian
parametrisation of the following
form \cite{WH96a} (and cf. \cite{WH96b}): 
\begin{equation}
\label{17}
C(q,K) - 1 = \exp  [- R_\perp^2({\bf K})q_\perp^2 -
   R_z^2({\bf K}) q_l^2 - R_t^2({\bf K})(q^0)^2 -
   2\, R_{zt}^2({\bf K}) q_l q^0 ] \, .
\end{equation}
This parametrisation might provide better insight to 
the sources of possible statistical uncertainties, since 
its parameters have clear geometric information concerning
the Gaussian shape of the correlation function in the
($q^0$, $q_l$)-plane. The 
YKP parameters could be then calculated from the parameters
(\ref{17}) by 
using the following formulae (for $R_{zt} \ne 0$):
\begin{eqnarray}
v & = & - \frac {R_z^2 + R_t^2 - \sqrt{{(R_z^2 + R_t^2)}^2
    - 4\, R_{zt}^2}}{2\, R_{zt}^2}\, ,
 \label{18a} \\
R_\parallel^2 & = & \frac 
    {R_z^2 - R_t^2 + \sqrt{{(R_z^2 + R_t^2)}^2
    - 4\, R_{zt}^2 }}2  = \frac {R_z^2 - v^2 R_t^2}
    {1 + v^2} \, ,
 \label{18b} \\
R_0^2 & = & \frac {R_t^2 - R_z^2 + \sqrt{{(R_z^2 + R_t^2)}^2
    - 4\, R_{zt}^2 }}2 = \frac {R_t^2 - v^2 R_z^2}
    {1 + v^2}\, ,
 \label{18c}
\end{eqnarray}
while $R_{zt} = 0 \Rightarrow v = 0$, 
$R_\parallel = R_z$ 
and $R_0 = R_t$.

\begin{figure}[ht]
\begin{center}
\begin{minipage}{12cm}
\epsfxsize=12cm
\epsfysize=11.2cm
\epsfbox{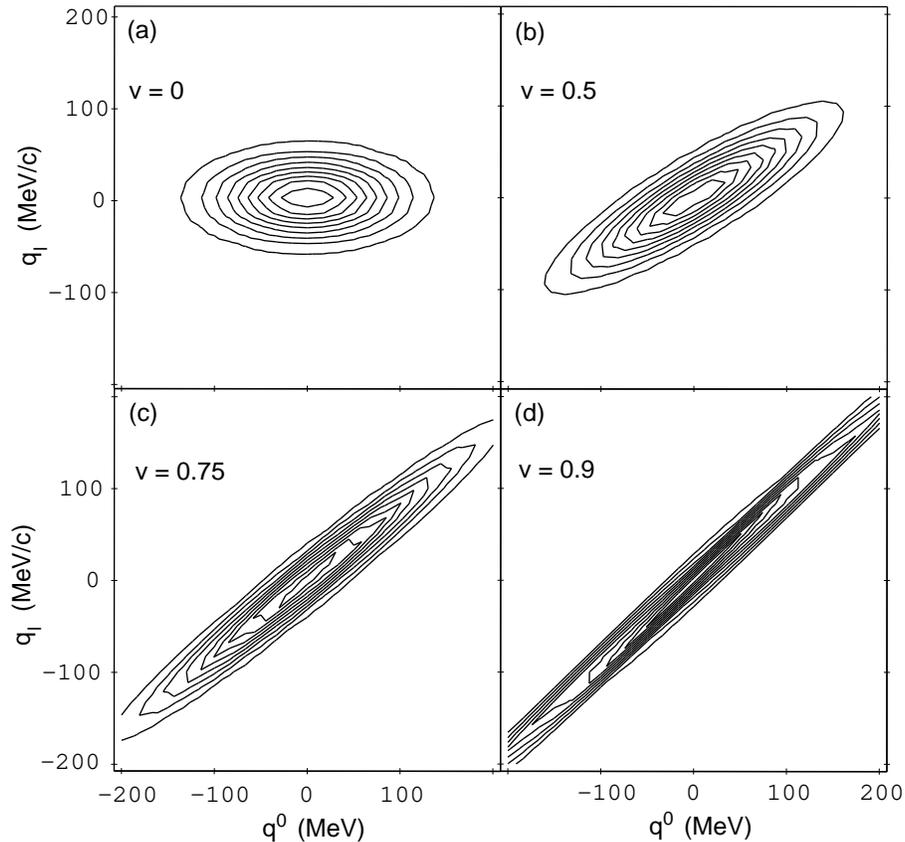}
\caption{Contour plots of the YKP parametrisation 
of the correlation function 
projected into the ($q_l$-$q^0$) plane. The values of the
radii are: $R_\parallel = 5$ fm, $R_0 = 2.2$ fm/c. The
values of the YK velocity on different plots are:
a) $v = 0$, b) $v = 0.5$, c) $v= 0.75$, d) $v=0.9$.} 
\label{f1}
\end{minipage}
\end{center}
\end{figure}

The second idea which could lead to the improvement of the
fitting is the choice of the appropriate reference frame.
Although the radii are longitudinally boost invariant, the 
shape of the correlation function is drastically changed 
when boosting to another reference frame. This is shown
on  Fig.~\ref{f1}. 
In the YK frame the projection of the
correlation function into the $q_l$-$q_0$ plane is not rotated
and has ``reasonable'' widths. When increasing the value of $v$
for fixed $R_\perp$, $R_\parallel$, $R_0$, 
i.e., when boosting to another frame, 
the projection rotates to the diagonal and becomes very
narrow. A shape of this kind may cause additional 
problems in the fitting procedure unless the binning
of the data is adjusted to
the shape of the correlation function.

\subsection{Model studies}
\label{model}

\begin{figure}[ht]
\begin{center}
\begin{minipage}{13cm}
\epsfxsize=13cm
\epsfysize=10cm
\epsfbox{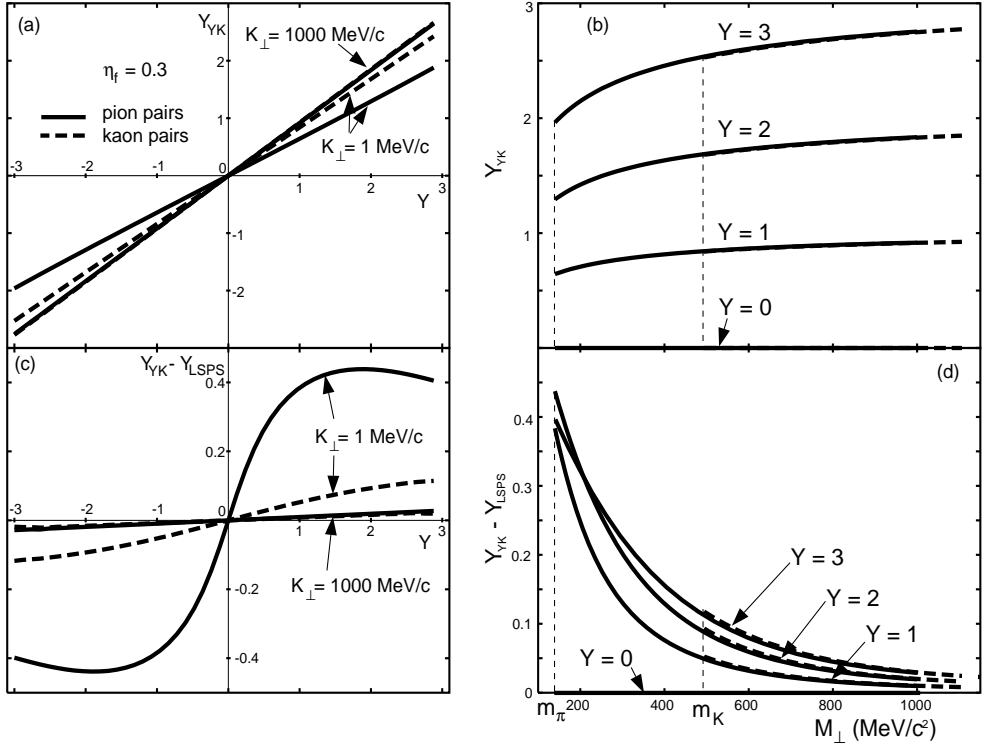}
\caption{The YK rapidity, defined as $Y_{_{\rm YK}} = 0.5 \,  
\ln (1+v)/(1-v)$. Transverse flow $\eta_f = 0.3$,
solid lines -- pions, dashed lines -- kaons. a) Dependence of 
$Y_{_{\rm YK}}$ on the rapidity of the pair $Y$ for fixed 
$K_\perp$; 
b) dependence of $Y_{_{\rm YK}}$ on $M_\perp$ for fixed $Y$;
c) dependence of the difference between $Y_{_{\rm YK}}$
and the rapidity of the point of maximal emissivity
$Y_{_{\rm LSPS}}$ on $Y$ for fixed $K_\perp$;
d) same as c) but dependence on $M_\perp$ for fixed $Y$.
} 
\label{f2}
\end{minipage}
\end{center}
\end{figure}
To illustrate the main properties of the YKP radii
we present here results obtained from a model 
study. The radii are calculated numerically using the
formulae (\ref{13a}--\ref{13c}) and 
(\ref{14a}--\ref{14d}). As a model for 
the source the emission function of ref.\cite{CNH95}
which is a special case of the emission function 
from ref.\cite{CL96} is taken. It is given by the formula
\begin{equation}
S(x,K) = \frac {M_\perp \cosh(\eta - Y)}
  {{(2\pi)}^3 \sqrt{2\pi{(\Delta \tau)}^2}} \exp \left [ -
  \frac{K\cdot u(x)}{T}\right ] \exp \left [ -
  \frac {r^2}{2R^2} - \frac {{(\eta - \eta_0)}^2}{2{(\Delta \eta)}^2}
  - \frac {{(\tau - \tau_0)}^2}{2{(\Delta \tau)}^2} \right ] \, .
\label{19}
\end{equation}
The first part of this emission function describes the
geometry of the freeze-out hypersurface. The second is
a Lorentz invariant 
Boltzmann distribution which reflects the assumption
of local thermal equilibrium at freeze-out. 
The last exponential describes
the finite geometrical size of the source
in the transverse direction,
space-time rapidity and  proper time. The motion of the different 
fluid elements of the source is expressed by the velocity field 
$u(x)$. Here we assume longitudinal expansion of 
the Bjorken type \cite{Bjork83} and a linear transverse 
expansion profile. 
Then the components of $u(x)$ are given by:
\begin{equation}
u(x) = (\cosh \eta \, \cosh \eta_t(r),\, 
   \sinh \eta_t(r) \frac xr ,\,
   \sinh \eta_t(r) \frac yr ,\, \sinh \eta \, \sinh \eta_t(r) )
\label{20}
\end{equation}
with
\begin{equation}
\eta = \frac 12 \ln \frac {t + z}{t - z}
\label{21}
\end{equation}
and
\begin{equation}
\eta_t(r) = \eta_f \frac rR
\label{22}
\end{equation}
The parameter $\eta_f$ scales the strength of the transverse 
flow.

The temperature $T$ in the calculation was set to 140 MeV,
the transverse geometric radius $R$ is 3 fm, the space-time 
rapidity width $\Delta \eta = 1.2$, the average freeze-out proper 
time
$\tau_0 = 3$ fm/c, and the mean proper emission duration
$\Delta \tau = 1$ fm/c. Calculations were done for pions
($m_\pi = 139$ MeV) and kaons ($m_K = 494$ MeV).

\begin{figure}[ht]
\begin{center}
\begin{minipage}{11cm}
\epsfxsize=11cm
\epsfysize=11.5cm
\epsfbox{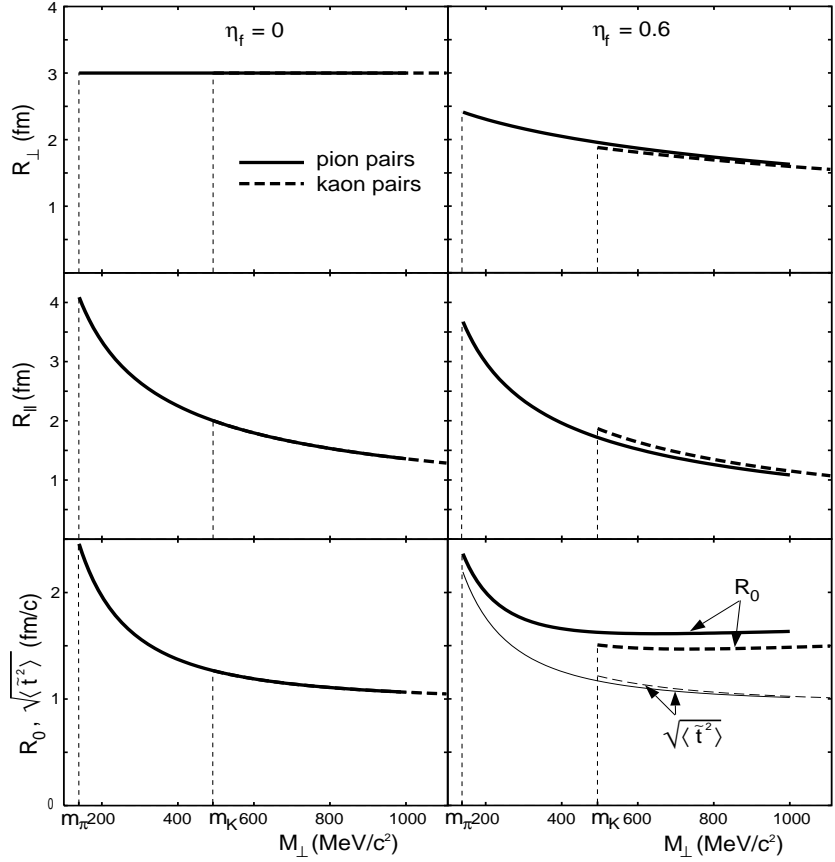}
\caption{Dependences   of the
YKP radius parameters on $M_\perp$. 
Solid lines -- pions, dashed lines -- kaons. 
Left column: $\eta_f = 0$,
right column: $\eta_f = 0.6$. 
The lines for $R_0$ in the case with 
the transverse flow are also 
compared to the real effective lifetime
${ \left \langle {\tilde t}^2 \right \rangle}^{\frac 12}$.
Without  transverse flow 
${\left \langle {\tilde t}^2 \right \rangle}^{\frac 12}$
coincides with $R_0$.
} 
\label{f3}
\end{minipage}
\end{center}
\end{figure}

The behavior of the YK velocity in this model is plotted
in Fig.~\ref{f2} \footnote{Figures \ref{f2} and \ref{f3}
are taken from Ref.\cite{HTWW96b}.}. 
The results are presented using the 
rapidities instead of velocities. The Yano-Koonin
rapidity is connected with the YK velocity via
the usual relation
\begin{equation}
\label{22a}
Y_{_{\rm YK}} = \frac 12 \ln \frac {1+v}{1-v}
\end{equation}
Fig.~\ref{f2}a shows the dependence of the 
YK rapidity on the rapidity of the pair. The linear 
increase is a consequence of the assumed longitudinal
expansion with a Bjorken flow profile. It is also seen
that the
lines corresponding to the pairs with higher 
transverse mass are closer to the diagonal. 
They reach the diagonal in the limit $M_\perp \to 
\infty$. This tendency is evident in Fig.~\ref{f2}b.
The YK rapidity is sensitive mainly to the longitudinal
rapidity of the pair, the dependence on its transverse
mass being weaker.  The $M_\perp$ dependences for pions
and kaons almost coincide; the slight breaking  of 
$M_\perp$ scaling is caused by the non-vanishing transverse
flow \cite{HTWW96b}. 

The plots c) and d) of Fig.~\ref{f2} show the difference 
between
the YK rapidity and the rapidity of the point of maximal
emissivity of the source for given pair rapidity and 
transverse mass.
This difference is appreciable only in the region
with non-zero pair rapidity and small $M_\perp$. But
already in the region of measurements with kaons the
difference is very small.

In Fig.~\ref{f3} the characteristic behavior of the YKP
radius parameters is plotted. 
The two columns correspond
to different physical scenarios: 
the left one is without transverse
flow while in the right one the value of $\eta_f$ is
quite big ($\eta_f=0.6$). 
Before embarking on an 
interpretation of the dependences let us recall
that the HBT radii measure the homogeneity lengths and
not the dimensions of the whole fireball.

It is clearly seen that the strong transverse flow
breaks the common $M_\perp$ scaling -- the lines for
pions and kaons do not coincide. Without the transverse
flow the only parameter in the emission function is
the transverse mass and the $M_\perp$ scaling is restored.
The influence of transverse flow
is most evident in the behavior of 
$R_\perp$. For $\eta_f = 0$ there is no dynamics
which could influence the transverse radius, hence
it is given by the geometric transverse radius of the source.
In the real experiment there could be another 
dynamical effects leading to the non-trivial $M_\perp$
dependence, e.g. temperature gradients. But the only
yet known effect, within the class of thermal models,
leading to the
breaking of common $M_\perp$ scaling is the transverse 
flow.

The decrease of $R_\parallel$ is a result of the
longitudinal dynamics with strong velocity gradient.
The fluid element with given $z$-coordinate moves
with the longitudinal velocity 
\begin{equation}
\label{23}
v = \frac zt \, .
\end{equation}
The particle with some longitudinal velocity is
effectively produced by the fluid element with the
same velocity and its surroundings with velocities
within the thermal smearing. The thermal smearing in this way
delimits for higher $M_\perp$ a smaller velocity interval
than for lower $M_\perp$. According to eq.(\ref{23}) the 
longitudinal radius decreases. Again,
transverse flow destroys the common $M_\perp$
scaling.

The behavior of $R_0$ is related to the $M_\perp$
dependence of $R_\parallel$. It is given by the
geometry of the freeze-out hypersurface. 
This source freezes out along the hyperbola in
z-t diagram. Thus events at the freeze-out hypersurface
with different longitudinal coordinates occur at 
different times and the longitudinal homogeneity length
determines partially the effective emission
duration (lifetime) \cite{CSH95b,CL96}. If the longitudinal
homogeneity length vanishes (i.e. at large $M_\perp$), 
the lifetime is given
by the geometric temporal term with width $\Delta \tau$.

In the case with  transverse flow, the difference
between $R_0$ and the effective emission duration is
plotted too. For high $M_\perp$, $R_0$ is about 50\%
bigger than the real lifetime of the source, but note,
that the assumed value of $\eta_f$ is rather high. 
For realistic cases we expect $\eta_f$ about 
0.2 - 0.3. If the transverse flow vanishes, $R_0$ measures
the lifetime exactly.

\section{Conclusions}
\label{conc}

The Yano-Koonin-Podgoretski\u\i \ parametrisation
of the two-particle correlation function has been
presented. Comparing to the older and
widely used Cartesian parametrisation it provides 
better insight
into the dynamics of the source. 

A great advantage is the direct measurement of the
emission duration provided by $R_0$. The lifetime
is the leading term in the model-independent 
expression for $R_0$ and it 
is measured with some small deviations arising from 
the non-vanishing transverse flow.

The fourth fit parameter, the YK velocity or YK
rapidity, offers a possibility to study the longitudinal
dynamics of the source. Its linear increase with
the rapidity of the pair is a signal for a longitudinal
expansion. On the other side, if $Y_{_{\rm YK}}$
does not depend on the pair rapidity, this is a signal for
a source without longitudinal expansion,
producing particles with  
all rapidities from a source at rest in the CM. 
Please note, that these statements are 
not based on model independent results but on the 
results of model studies within a class of thermal
models. 

{\bf Acknowledgements: }  
We thank  H. Appelsh\"auser,
D. Ferenc, K. Kadija, J. Pi\v{s}\'ut and C. Slotta 
for stimulating and clarifying discussions. B.T. expresses
his thanks to the organizers of the Workshop on
Heavy Ion Collisions in Bratislava for the invitation
and to the Dept. of Theor. Physics of Comenius
University for kind hospitality. This work was supported by DAAD,
DFG, NSFC, BMBF and GSI.


\end{document}